\documentclass[twocolumn,showpacs,preprintnumbers,amsmath,amssymb]{revtex4}

\usepackage{graphicx}% Include figure files
\usepackage{dcolumn}% Align table columns on decimal point
\usepackage{bm}% bold math

\def\lapproxeq{\lower .7ex\hbox{$\;\stackrel{\textstyle
<}{\sim}\;$}}
\def\gapproxeq{\lower .7ex\hbox{$\;\stackrel{\textstyle
>}{\sim}\;$}}

\begin{document}

\hyphenation{Karls-ru-he}

% \preprint{DRAFT!!!}

\title{Geometric Structures in Hadronic
Cores of Extensive Air Showers Observed by KASCADE}

\author{
T.~Antoni$^1$,
W.D.~Apel$^2$,
F.~Badea$^{2}$
}
\altaffiliation[On leave of absence from ]{(3)}
\author{
K.~Bekk$^2$,
A.~Bercuci$^{3}$,
% }
% \altaffiliation[On leave of absence from ]{(3)}
% \author{
J.~Bl\"umer$^{2,1}$,
H.~Bozdog$^2$,
I.M.~Brancus$^3$,
% C.~B\"uttner$^1$,
A.~Chilingarian$^4$,
K.~Daumiller$^2$,
P.~Doll$^2$,
R.~Engel$^2$,
J.~Engler$^2$,
F.~Fe{\ss}ler$^2$,
H.J.~Gils$^2$,
R.~Glasstetter$^{1}$
}
\altaffiliation[Now at ]{University of Wuppertal, 42117 Wuppertal, Germany}
\author{
A.~Haungs$^2$,
D.~Heck$^2$,
J.R.~H\"orandel$^1$,
K-H.~Kampert$^{1,2}$
}
\altaffiliation[Now at ]{University of Wuppertal, 42117 Wuppertal, Germany}
\author{
H.O.~Klages$^2$,
G.~Maier$^2$
}
\altaffiliation[Now at ]{University of Leeds, LS2 9JT Leeds, UK}
\author{
H.J.~Mathes$^2$,
H.J.~Mayer$^2$,
J.~Milke$^2$,
M.~M\"uller$^2$,
R.~Obenland$^2$,
J.~Oehlschl\"ager$^2$,
S.~Ostapchenko$^{2}$
}
\altaffiliation[On leave of absence from ]{Moscow State University, 119899 Moscow, Russia}
\author{
M.~Petcu$^3$,
H.~Rebel$^2$,
A.~Risse$^5$
}
\altaffiliation[Corresponding author. ] {Electronic address: 
iwan@zpk.u.lodz.pl}
\author{
M.~Risse$^2$}
\author{
M.~Roth$^1$}
\author{
G.~Schatz$^2$}
\author{
H.~Schieler$^2$}
\author{
J.~Scholz$^2$}
\author{
T.~Thouw$^2$}
\author{
H.~Ulrich$^2$}
\author{
J.~van~Buren$^2$}
\author{
A.~Vardanyan$^4$}
\author{
A.~Weindl$^2$}
\author{
J.~Wochele$^2$}
\author{
J.~Zabierowski$^5$
\vspace{0.5cm}
}
% \email{alvarez@bartol.udel.edu}
\affiliation{
(1) Institut f\"ur Experimentelle Kernphysik, Universit\"at Karlsruhe, 76021~Karlsruhe, Germany
\\
(2) Forschungszentrum Karlsruhe, Institut f\"ur Kernphysik, 76021~Karlsruhe, Germany
\\
(3) National Institute of Physics and Nuclear Engineering, 7690~Bucharest, Romania
\\
(4) Cosmic Ray Division, Yerevan Physics Institute, Yerevan~36, Armenia
\\
(5) Soltan Institute for Nuclear Studies, 90950~Lodz, Poland
}

% \date{\bf DRAFT \today}
% \date{\bf DRAFT: 9 August 2004}

\begin{abstract}

The geometric distribution
of high-energy hadrons $\ge$100~GeV in shower cores
measured with the KASCADE calorimeter is analyzed.
The data are checked for sensitivity to hadronic interaction
features and indications of new physics as discussed in the
literature.
The angular correlation of the most energetic hadrons and 
in particular the fraction of events with hadrons being aligned 
are quantified by means of the commonly used parameter $\lambda_4$. 
The analysis shows that the observed $\lambda_4$ distribution is
compatible with that predicted by simulations and
is not linked to an 
angular correlation from hadronic jet production at high energy.
Another parameter, $d_{4}^{\rm max}$, describing distances between 
hadrons measured in the detector, is found to be sensitive both to the
transverse momenta in secondary hadron production and the primary particle
type.
Transverse momenta in high-energy hadron interactions differing by
a factor two or more from what is assumed in the standard simulations are
disfavoured by the measured $d_{4}^{\rm max}$ distribution.

\end{abstract}

% PACS meaning
% 96.40.Pq Extensive air showers
% 96.40.-z Cosmic rays
% 13.85.-t Hadron-induced high- and super-high-energy interactions
% 13.85.Tp  Cosmic-ray interactions


\maketitle

\section{\label{introduction}Introduction}

High-energy hadrons in extensive air showers offer a unique
possibility to study interaction features well beyond the kinematic
and energy region of earthbound accelerators.
In particular, structures in hadronic shower cores might
reflect properties of the particle production at an initial stage of the
shower development.  For example,
the EAS-TOP Collaboration investigated multicore events recorded in the
calorimeter and studied the cross-section of large $p_t$ jet
production in proton-air collisions~\cite{eastop}.
In other studies,
data taken by the KASCADE scintillator array and
hadron calorimeter~\cite{kascade,calor} were used to 
investigate different aspects of hadronic interaction models. 
The correlation of the hadronic
to the muonic and electromagnetic shower components~\cite{had1,milke},
as well as features of the particle production in the very forward region
of nucleon-air collisions~\cite{had2} were measured and compared to model
predictions.

Geometric structures in hadronic shower cores
at observation level are particularly
interesting, as one expects QCD jet production to lead to 
secondary hadrons being naturally aligned to form line shape
patterns~\cite{halzen}. Similar alignment structures might result
from exotic hadron production processes \cite{roizen}.
Therefore, it is not surprising that 
the observation of aligned structures in a number of 
emulsion chamber experiments has initiated
considerable experimental and theoretical efforts
over the last two decades~\cite{defl4,kop95,pamir,bor97,otherconf,
highpt,highpt2,capsla99,na22,runjob,kanbala,strato,cap01,halzen,
roizen,mukh,troshin,white,rest,icrc03}.

Aligned event structures were, for instance, reported by the
PAMIR experiment~\cite{pamir,kop95}.
An excess of events with substructure alignment  
above background fluctuations
was found in the data for primary energies above
an energy threshold of 8$-$10~PeV 
(see e.g.~\cite{kop95,bor97,otherconf,highpt,highpt2,capsla99}
and references therein).
The most energetic event of $\simeq$10~PeV that was measured
in an emulsion experiment 
during flights with the Concorde airplane also shows alignment~\cite{cap01}.
Furthermore, alignment was found in another individual high-energy event, 
estimated to have an energy $>$10~PeV, that was detected
by a balloon borne emulsion chamber~\cite{strato}.

The experimental evidence of the existence of an alignment phenomenon,
i.e.~events with substructures being located in a line
occurring more frequently than expected, is
debated controversially, including the possibility of the excess being
due to statistical fluctuations (see e.g.~\cite{cap01}).

At lower energy, no excess of aligned structures in air showers or
direct experiments was observed.
At energies of a few PeV the fraction of elongated
events measured by PAMIR was found to agree with the expectations from
background fluctuations.
The fraction of elongated events of similar energy recorded by the 
Kanbala air shower experiment~\cite{kanbala} gave also no indication of
a significant excess over the simulated background. 
Measurements of the balloon experiment RUNJOB
in a primary energy range of 0.01$-$0.1~PeV have shown that
the fraction of aligned events agreed with the
background expectation~\cite{runjob}.
Also the less energetic events recorded during the Concorde flights
showed no excess alignment.

A search for such alignment phenomena at accelerators was performed at 
comparatively low collision energy of 250~GeV in $\pi$$-$Au interactions. 
No significant excess of elongated events was found in data of the
CERN experiment NA22~\cite{kop95,na22}.

Various theoretical efforts, partly hampered by the limited
experimental guidance,
were performed to explain alignment by physics mechanisms.
It was pointed out in~\cite{halzen} that
dynamical features of standard QCD jet production might give rise
to observable alignment.
In~\cite{kop95,capsla99} it was emphasized that the observations would
indicate aligned particle production in the fragmentation region of
the collision rather than in the central region.
A connection of alignment to the production of secondaries
with high transverse momentum was suggested
e.g. in~\cite{highpt,highpt2}.
The appearance of an anti-shadow scattering mode at high collision
energies as source of alignment was discussed in~\cite{troshin}.
An onset of semi-hard double inelastic diffraction was
proposed in~\cite{roizen}, where alignment results as projection
from the rupture of a quark-gluon string.
The existence of a new particle
with a long mean free path, produced in a new type of interaction,
was also proposed to explain the PAMIR 
observation~\cite{capsla99,white}.

Simulations based on phenomenological models were compared
to the PAMIR data in~\cite{highpt2,mukh}.
It was pointed out that alignment features produced, e.g., in the first
interaction of the primary cosmic ray in air can be washed out by
subsequent interactions during the shower evolution. 

While the existence of alignment features is out of doubt, the main
questions concerning their nature are: To what extent can
such structures observed in the data be related to specific particle
production and interaction mechanisms?
Are new particles or interaction channels, maybe above a certain
collision energy threshold, required to explain the data?

In this work, geometric structures in hadronic shower cores recorded
by the KASCADE experiment are investigated. Both, 
the alignment of hadrons in shower cores and the distance between
hadrons are analyzed and compared to detailed Monte Carlo simulations.

The article is organized as follows.
In Section~\ref{cuts}, relevant features of 
the KASCADE detector and the applied data
selection are described.
Results on aligned geometric structures are presented in 
Sec.~\ref{results}, including a detailed
study of the background expectation due to random fluctuations.
Another observable, characterizing the distance 
between hadrons, is introduced in Sec.~\ref{d4max} and used to 
analyze KASCADE data.  Implications and 
comparisons to results from other experiments are
discussed in Sec.~\ref{conclusions}.

%%%%%%%%%%%%%%%%%%%%%%%%%%%%%%%%%%%%%%%%%%%%%%%%%%%%%%%%%%%%%%

\section{\label{cuts} Measurements }

The KASCADE air shower experiment~\cite{kascade} is
located at an altitude
of 110~m a.s.l.~(corresponding to 1020~g~cm$^{-2}$ atmospheric depth)
at Karlsruhe, Germany. Its 16$\times$20~m$^2$
hadron calorimeter~\cite{calor} is well suited to 
investigate geometric structures in hadronic shower cores.
Individual hadrons with energies from 50~GeV up to more than 20~TeV
are measured with a spatial resolution of $\simeq$15~cm.
The primary energy covered by KASCADE includes the 
range of 8$-$10~PeV claimed as threshold energy for alignment.
A detailed description of the KASCADE detector performance and 
the standard reconstruction procedures for the array and calorimeter data
is given in~\cite{kascade,calor}.

For the analysis, data collected by the KASCADE experiment in the
period from May 1998 to April 2001 have been used.
Primary energy, direction, and core position of air showers with 
energies $>$0.5~PeV are determined with the scintillator array
data.
The energy, position and direction of individual hadrons with energies
$\ge$50~GeV are reconstructed from the measurements performed with
the hadron calorimeter.
The following selection criteria are applied to the data:\\
(i) The reconstructed primary energy of the air shower
($\Delta E_0/E_0$ $\simeq$25\%)
 is $\ge$1~PeV, and the primary zenith angle is 
 $\le$30$^\circ$ ($\Delta \theta$ $\simeq$1$^\circ$).\\
(ii) The reconstructed shower core location
($\Delta x_c$ $\simeq$1.5~m) is contained in the active calorimeter
area with a minimum distance to the calorimeter boundaries of 3~m.
\\
(iii) At least four hadrons with energies $\ge$100~GeV are reconstructed
($\Delta E_h/E_h$ $\simeq$20\%).\\
The effect of the latter condition is small even at lower primary
energies, reducing the data set by $\simeq$13\%.

\begin{figure}
\centerline{
\includegraphics[width=7.36cm]{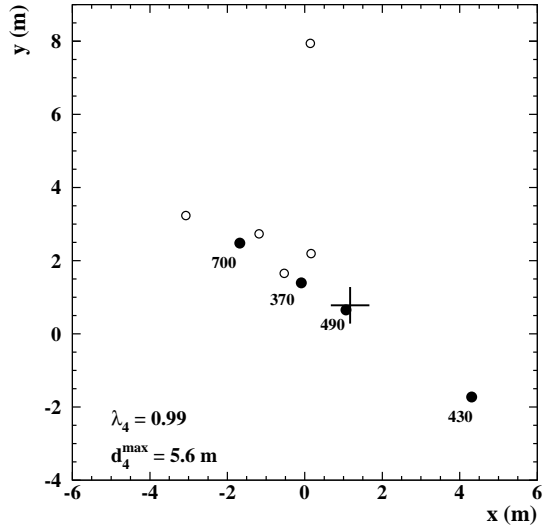}
}
\caption{Example of a measured hadronic shower core 
(hadron positions in the shower plane)
with $\lambda_4 = 0.99$ and $d_{4}^{\rm max} = 5.6$~m.
9 hadrons with energies above 100~GeV are reconstructed.
For the four most energetic hadrons (full symbols), the 
energies in GeV are given.
The shower core position as reconstructed by
the scintillator array is marked by a cross.
The active calorimeter area exceeds the area plotted.}
\label{fig-bsp}
\end{figure}

In total, 4489 events are selected this way.
After transformation into the shower plane, the quantities
$\lambda_4$ and $d_{4}^{\rm max}$ (see below) are evaluated
for each event. 
An example of a measured elongated event is given in
Fig.~\ref{fig-bsp}.
%%%%%%%%%%%%%%%%%%%%%%%%%
%%%%%%%%%%%%%%%%%%%%%%%%%

%%%%%%%%%%%%%%%%%%%%%%%%%%%%%%%%%%%%%%%%%%%%%%%%%%%%%%%%%

\section{\label{results} Alignment of hadrons}

Commonly, the degree of alignment is described by a parameter
$\lambda_4$~\cite{defl4}
quantifying the angular correlation between the four most-energetic
particles (or particle clusters):
\begin{equation}
\label{eq-l4}
\lambda_4 = \frac{1}{24}\cdot\sum^4_{i\neq j\neq k}
\cos 2\varphi^k_{ij} ~~,
\end{equation}
where $\varphi^k_{ij}$ denotes the angle between the lines
connecting particle $k$ to $i$ and $j$.
Possible values range between $\lambda_4=-\frac{1}{3}$
(isotropic distribution) and $\lambda_4 = 1$ (perfect alignment).
Events are usually termed ``aligned'' or ``elongated'' for $\lambda_4 \ge 0.8$.

\begin{figure}
\centerline{
\includegraphics[width=8.0cm]{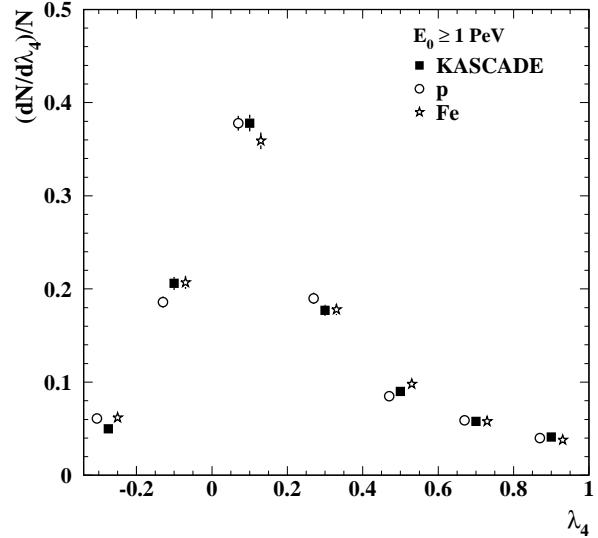}
}
\caption{$\lambda_4$ distribution measured by KASCADE compared to
simulation results for primary proton and iron showers.
For clarity, simulation points are slightly displaced horizontally.}
\label{fig-kco}
\end{figure}

The measured $\lambda_4$ distribution is displayed in Fig.~\ref{fig-kco}.
Also shown are the results for primary proton and iron nuclei 
simulated with the CORSIKA (v6.0) code~\cite{corsika} 
employing the QGSJET~01~\cite{qgsjet01} hadronic interaction model. 
The simulations are performed for a continuous primary
energy spectrum with a spectral index $-2.7$.
It has been checked explicitely that adopting other values for the
primary energy slope or, e.g., including a {\it knee} structure
does not influence the calculated $\lambda_4$ distribution, as will
be also clear from the following discussion.
The shower simulation is followed by a detailed detector simulation
based on GEANT~\cite{geant}. The same reconstruction algorithms and
selection criteria are applied to simulations and data.
The simulation statistics is of comparable size to the
data statistics for each primary.

Fig.~\ref{fig-kco} shows that the measured $\lambda_4$ distribution
and the fraction of aligned events are well reproduced by the
standard calculations, without the need of introducing new particles
or non-standard interaction features.
Moreover, despite the much smaller initial energy per nucleon in case
of iron
showers, no significant difference between proton and iron induced 
events is observed. This already indicates a minor sensitivity of the
observed parameter $\lambda_4$ to hadronic particle production in the
first interactions.
%%%%%%%%%%%%%%%%%%%%%%%%%%%%%%%
%%%%%%%%%%%%%%%%%%%%%%%%%%%%%%%

\begin{figure}
\centerline{
\includegraphics[width=8.0cm]{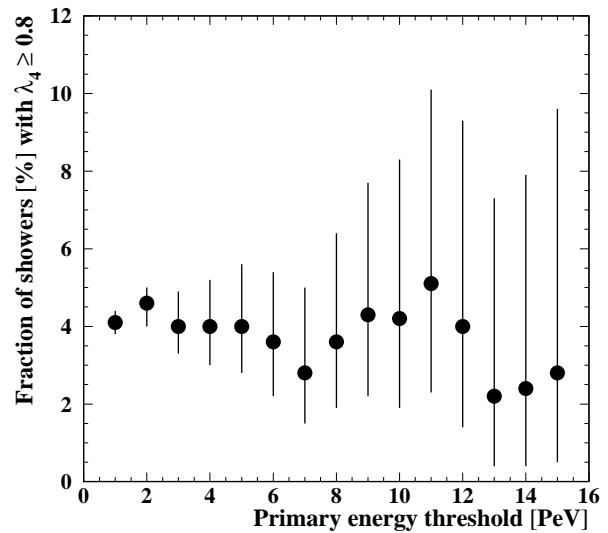}
}
\caption{Fraction of aligned events with $\lambda_4 \ge 0.8$
as measured by KASCADE versus the primary energy threshold.}
\label{fig-ethresh}
\end{figure}

The measured fraction of aligned events ($\lambda_4 \ge 0.8$) is plotted
in Fig.~\ref{fig-ethresh} as a function of the primary energy threshold
for selecting showers.
Within the statistical uncertainties, no energy dependence of the
fraction of elongated events is observed.
In particular, no significant increase or threshold behaviour for 
higher primary energies is found.
%%%%%%%%%%%%%%%%%%%%%%%%%%%%%%%
%%%%%%%%%%%%%%%%%%%%%%%%%%%%%%%

In the following, the sensitivity of
the observed $\lambda_4$ distribution and the fraction of
aligned event structures
to specific particle production features is investigated.
Events with aligned hadronic structures could be the result of jet
production in hadronic interactions during the shower
development~\cite{halzen}.
The jet pair and the hadrons from projectile fragmentation are
naturally aligned in one plane and the expected geometric 
separation falls in the range measurable with KASCADE.
For example, the projected separation from the initial direction
of a 10~TeV~hadron produced at an altitude of 15~km with a
transverse momentum $p_t = 5$~GeV 
is 7.5~m at KASCADE observation level.

\begin{figure}
\centerline{
\includegraphics[width=8.0cm]{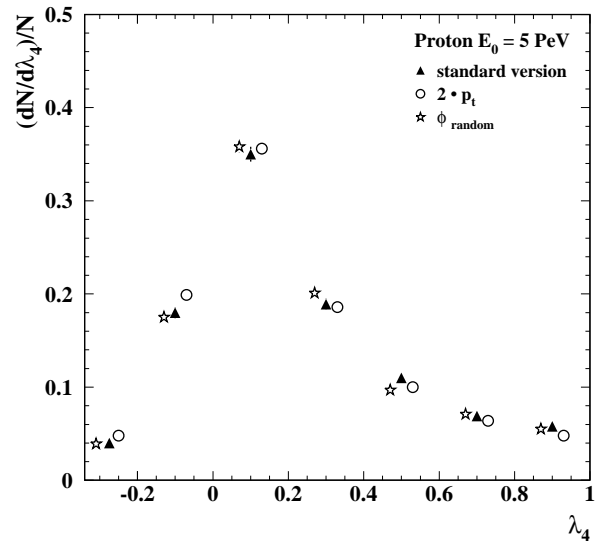}
}
\caption{$\lambda_4$ distributions for proton showers
of 5~PeV primary energy without modification of the
hadronic particle production (``standard version''),
with randomly chosen azimuth angles of secondary hadrons
(``$\Phi_{\mbox{random}}$''), and with transverse momenta
of secondary hadrons increased by a factor two (``$2\cdot p_t$'').
Simulation points obtained after the modifications
are slightly displaced horizontally.}
\label{fig-prot}
\end{figure}

To investigate the correlation between the angular distribution of
particles in hadronic interactions and geometric
structures, a simulation was carried out in which
the azimuth angles of the produced
secondary particles are picked at random in the centre-of-mass system of
the collision, keeping the transverse and longitudinal momenta
unchanged. In this way, any angular correlation of produced particles 
is eliminated.
The results of calculations performed for 5~PeV primary protons are shown
in Fig.~\ref{fig-prot} together with the standard simulations.
No significant change of the distribution as a whole
or the fraction of aligned events is observed for the
modified version (labeled ``$\Phi_{\mbox{random}}$'') with respect to the
original one.

Next, 
the sensitivity of $\lambda_4$ to the transverse momentum
$p_t$ of the secondaries is checked.
A connection of high-$p_t$ events and alignment was suggested,
e.g., in~\cite{highpt,highpt2}.
In the simulation, the $p_t$ of each secondary
produced in a hadronic interaction is artificially increased
by a factor two
(labeled ``$2\cdot p_t$'' in Fig.~\ref{fig-prot}), keeping the original
angular correlations unchanged. 
Such an increase is not realistic, but used here to study the sensitivity.
The overall shape of the $\lambda_4$ distribution is
affected very little.
No significant change of the fraction of aligned events can be noted.

%%%%%%%%%%%%%%%%%%%%%%%%%%%%%%%%
\begin{figure}
\centerline{
\includegraphics[width=8.0cm]{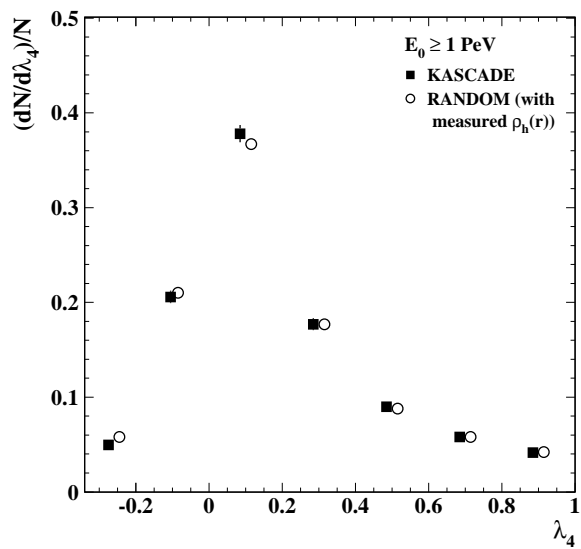}
}
\caption{$\lambda_4$ distributions: KASCADE data and
results from a simplified simulation of event topologies,
in which the azimuth angles were chosen randomly
with respect to the shower core,
and the distances according to the
measured hadron lateral distribution (see text).
Simulation points are slightly displaced horizontally.}
\label{fig-kra}
\end{figure}

%%%%%%%%%%%%%%%%%%%%%%%%%%%%%%%%
This insensitivity to modifications of the hadronic particle
production suggests that the measured
$\lambda_4$ distribution is mostly due to a random distribution
of the hadron azimuth angles.
Thus, a simplified simulation is performed by artificially generating
sets of four hadron positions in the shower plane.
The azimuth angles of the hadrons are chosen randomly with respect 
to the shower core. 
The lateral distances of the hadrons are sampled according to the
measured lateral distribution of the four most energetic hadrons.
The KASCADE data are indeed
reproduced, see Fig.~\ref{fig-kra}.
%%%%%%%%%%%%%%%%%%%%%%%%%%%%%%%%
%%%%%%%%%%%%%%%%%%%%%%%%%%%%%%%%

\begin{figure}
\centerline{
\includegraphics[width=8.0cm]{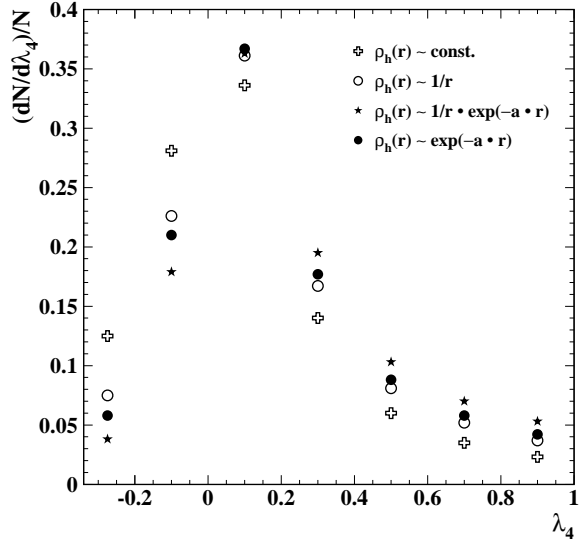}
}
\caption{$\lambda_4$ distributions obtained by 
a simplified simulation of event topologies (see text or
caption of Fig.~\ref{fig-kra}) for different hypothetical shapes of
the lateral distribution as indicated.}
\label{fig-lat}
\end{figure}

The sensitivity of the $\lambda_4$ distribution to the shape of the
lateral distribution of the observed particles
is checked with the simplified simulation
by varying the assumed lateral shape,
keeping the choice of hadron azimuth angles at random.
It should be noted that $\lambda_4$ as an observable,
being determined from the event topology,
does not depend on the absolute length scale of the assumed lateral
distribution, but only on its shape.
As can be seen in Fig.~\ref{fig-lat}, even choices of functional forms
that are unrealistic for air shower lateral distribution such as adopting
a particle density that is constant for any distance from the core,
result only in modest changes of the $\lambda_4$ distribution.
In general, a flatter lateral distribution
tends to decrease the fraction of aligned events generated in the random
sample.
However, the sensitivity of $\lambda_4$ to different lateral shapes 
is minor.
It seems hardly
possible for an air shower experiment to observe differences between
$\lambda_4$ distributions that are due to different lateral
distributions of the measured particles only.

%%%%%%%%%%%%%%%%%%%%%%%%%%%%%%%%%%%%%
%%%%%%%%%%%%%%%%%%%%%%%%%%%%%%%%%%%%%
Other parameters describing geometric structures similar to
$\lambda_4$ have been checked for sensitivity
to primary mass or hadronic interaction characteristics~\cite{thesis}.
For instance, all hadrons $\ge$100~GeV instead of the four most
energetic ones are considered to calculate a generalized quantity
$\lambda_n$. Also, weight factors for the hadrons are introduced
to additionally take their reconstructed energy into account.
No significant sensitivity to hadronic particle production properties
of these $\lambda$-related quantities is found, either.

%%%%%%%%%%%%%%%%%%%%%%%%%%%%%%%%%%%%%%%%%%%%%%%%%%%%%%%%%%%%%%%%%%%

\section{\label{d4max} Distance between hadrons}

%%%%%%%%%%%%%%%%%%%%%%%%%%%%%%%%%%%%%%%%%

Hadronic shower observables with sensitivity to
interaction features have for instance been discussed
in~\cite{had1,had2}.
A specific, further observable that is based on the four most energetic
hadrons in each event as is $\lambda_4$, is given by the
maximum distance, $d_{4}^{\rm max}$,
between one of the four considered hadrons
to the geometric centre of the other three~\cite{icrc03}.
Since this quantity is intimately connected to the hadron lateral
distribution, a sensitivity both to hadronic interaction features such
as the transverse momentum of secondary hadrons and to
the primary particle type can be expected.

%%%%%%%%%%%%%%%%%%%%%%%%%%%%%%%%%%%%%%%%%%%%%
\begin{figure}
\centerline{
\includegraphics[width=8.0cm]{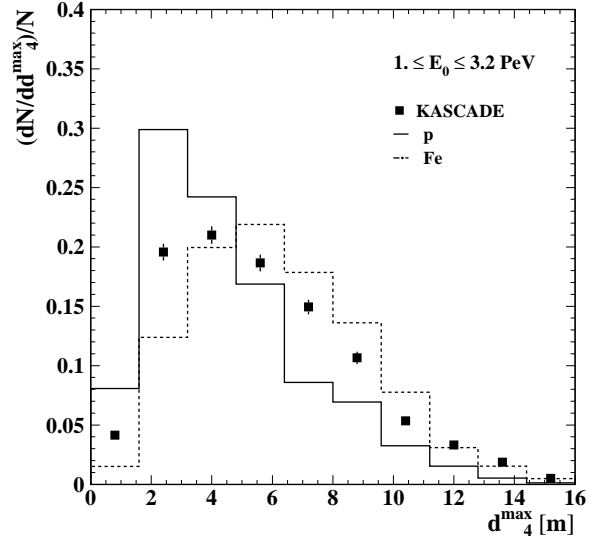}
}
\caption{
$d_4^{\rm max}$ distribution measured by KASCADE compared to simulation
results for primary proton and iron showers for primary energies of
1-3.2~PeV.
}
\label{fig-d4max}
\end{figure}
\begin{figure}
\centerline{
\includegraphics[width=8.0cm]{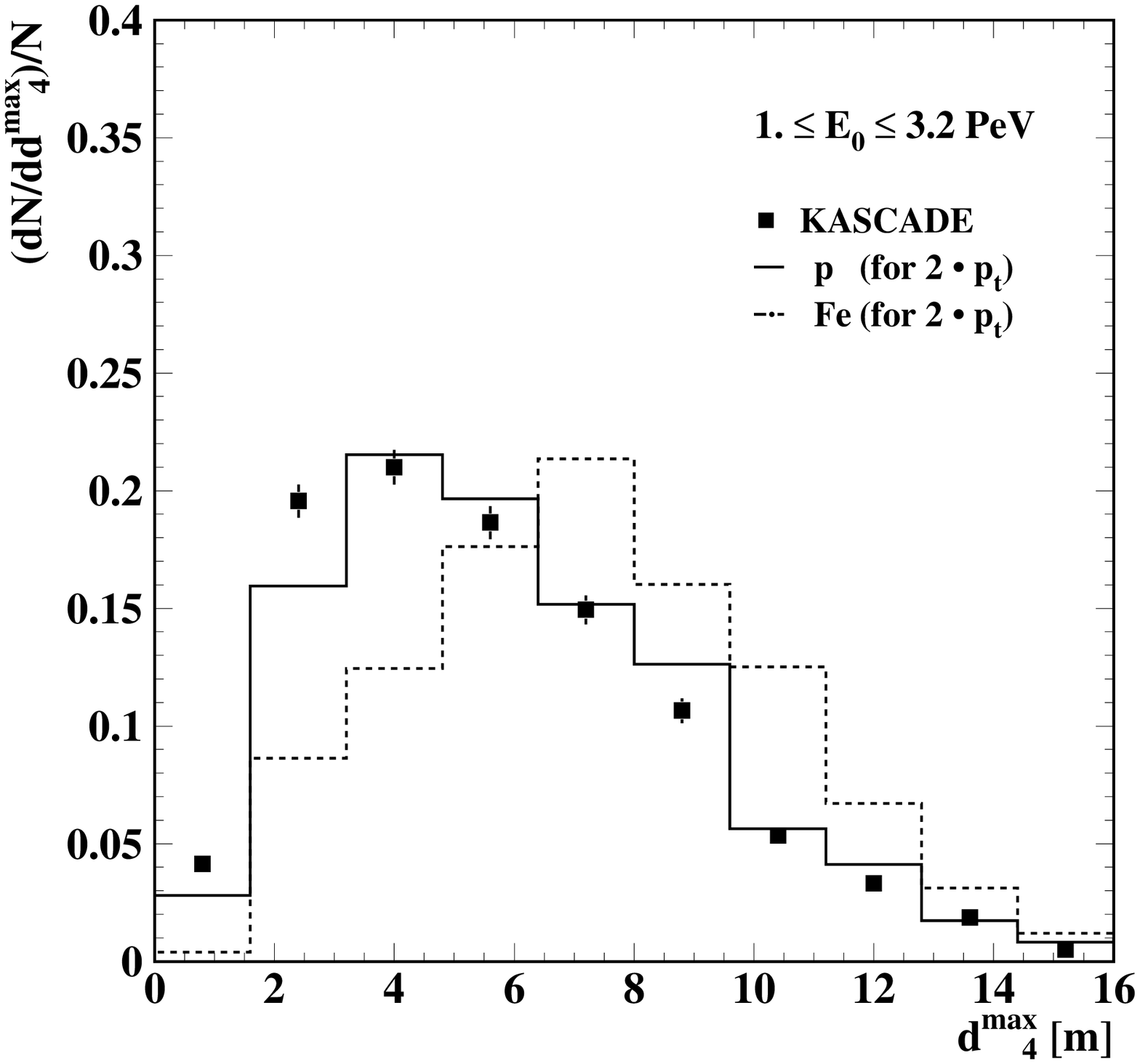}
}
\centerline{
\includegraphics[width=8.0cm]{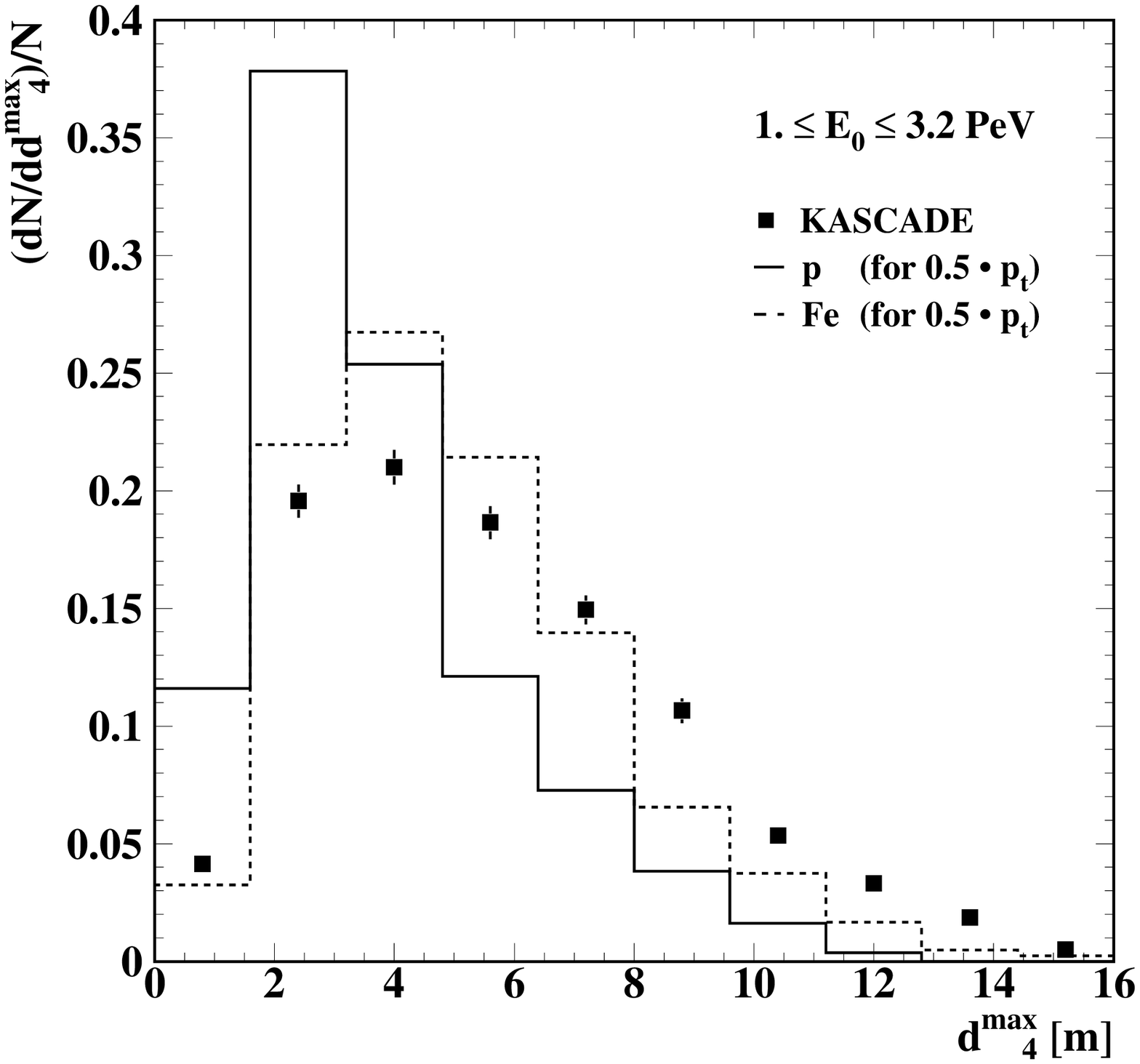}
}
\caption{
$d_4^{\rm max}$ distribution measured by KASCADE 
(see Fig.~\ref{fig-d4max}) compared to simulation
results for primary proton and iron showers with modified transverse
momentum of secondary hadrons. 
The transverse momenta were artificially increased 
(upper panel, ``$2\cdot p_t$'') and reduced 
(lower panel, ``$0.5\cdot p_t$'') by a factor two.
}
\label{fig-d4max2}
\end{figure}

%%%%%%%%%%%%%%%%%%%%%%%%%%%%%%%%%%%%%%%%%%%%%

The $d_{4}^{\rm max}$ distribution measured by
KASCADE at energies 1-3.2~PeV is compared to simulation results
for primary proton and iron showers in Fig.~\ref{fig-d4max}.
The comparison is restricted to energies below the {\it knee}
due to the increasing uncertainty of 
primary flux spectra and composition at higher energy.
The $d_{4}^{\rm max}$ distribution for primary protons is peaked at
smaller values compared to the predictions for iron-initiated events,
reflecting a steeper hadron lateral distribution in case of air showers
initiated by proton primaries.

The KASCADE data are mostly bracketed by
the primary proton and iron expectations. This seems reasonable,
as for these primary energies a mixed composition is favoured from
extrapolations of direct cosmic-ray measurements, from other air shower
experiments and in particular from the analysis of independent KASCADE
observables such as the electron and muon shower sizes~\cite{ulrich}.
Only at largest $d_{4}^{\rm max}$ values, a slight underestimation
shows up in the predictions that may indicate shortcomings in the
modelling of hadron interactions.

In the next step, as before,
the $p_t$ of secondary hadrons is artificially modified
in the simulations. The results are compared to KASCADE data in
Fig.~\ref{fig-d4max2}.
Assuming transverse momenta of secondary hadrons produced in
high-energy interactions twice as large as in standard simulations
(upper panel in Fig.~\ref{fig-d4max2}),
the distributions both for proton and iron primaries are significantly
shifted to larger $d_{4}^{\rm max}$ values. 
In this scenario, the simulations are hardly able to provide a 
satisfactory description of the $d_{4}^{\rm max}$ data. 
For any assumption on the primary composition between proton and iron
nuclei, and in particular for
a mixed composition as favoured by other measurements,
events with small values of $d_{4}^{\rm max}$ observed in hadronic
shower cores
are not adequately reproduced by simulations with doubled $p_t$.
The same conclusion holds when artificially
reducing the $p_t$ of secondary hadrons by a factor two
(lower panel in Fig.~\ref{fig-d4max2}).
In this case, the simulations fail to describe the events observed with
large values of $d_{4}^{\rm max}$.
Therefore,
hypothetical transverse momenta in high-energy secondary hadron production
that differ by a factor two or more from the standard assumptions
are disfavoured by the KASCADE data.

Another application of the $d_{4}^{\rm max}$ parameter is related to
investigating the primary cosmic-ray composition by means of
air shower simulations.
Due to the sensitivity to the primary particle type,
quantities such as $d_{4}^{\rm max}$ can be used to check the internal
consistency of hadronic interaction models.

%%%%%%%%%%%%%%%%%%%%%%%%%%%%%%%%%%%%%%%%%%%%%%%%%%%%%%%%%%%%%%%%%%%

\section{\label{conclusions} Discussion}

The results of the KASCADE data analysis concerning the $\lambda_4$
parameter can be summarized as follows.
The measured $\lambda_4$ distribution and the fraction of aligned events
of high-energy hadrons are well reproduced by
the simulations. 
Moreover, the data follow the expectations from randomly distributed
hadron azimuth angles.
No significant dependence on the primary energy was observed.
Within the statistical uncertainty, no correlation between $\lambda_4$
and hadronic interaction features such as jet production was found.

%%%%%%%%%%%%%%%%%%%%%%%%%%%%%%%%%
\begin{figure}
\centerline{
\includegraphics[width=8.0cm]{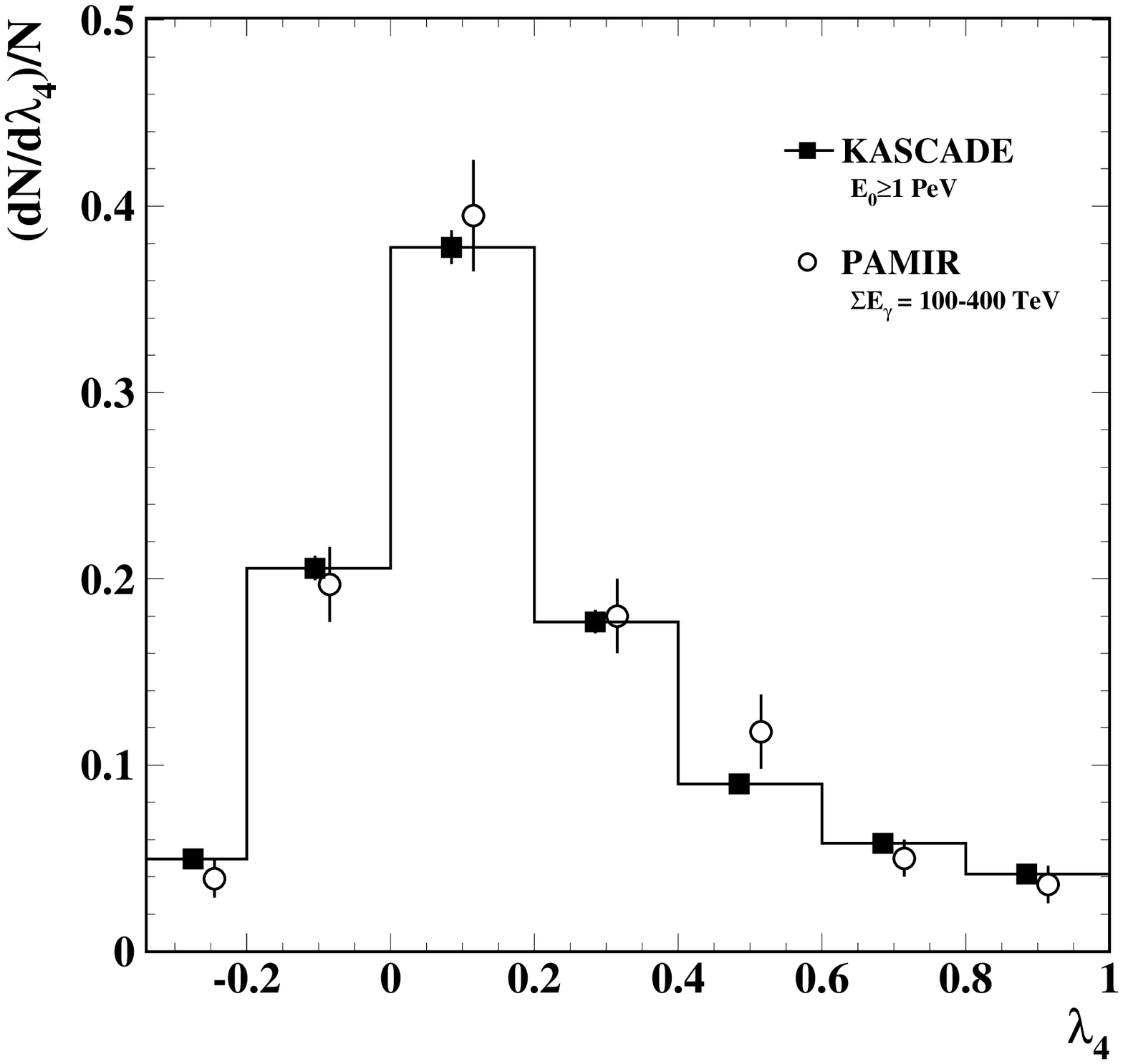}
}
\centerline{
\includegraphics[width=8.0cm]{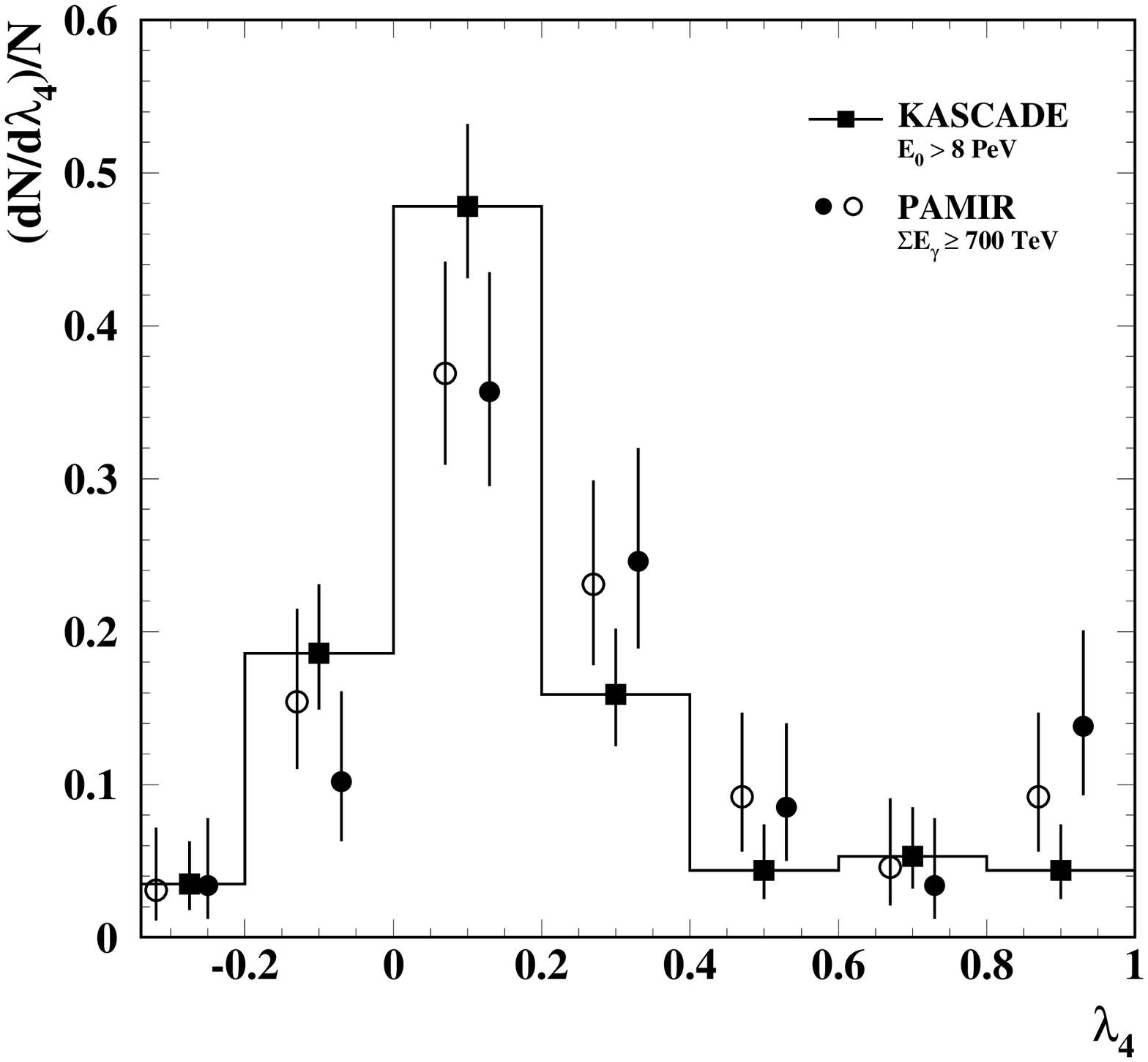}
}
\caption{$\lambda_4$ distributions: KASCADE and
PAMIR~\cite{capsla99,bor97} data. In the lower panel, the highest primary
energies are selected. The two distributions displayed for PAMIR
in the lower panel refer to different reconstructions 
from the same PAMIR data set~\cite{capsla99,bor97}.
PAMIR data are slightly displaced horizontally.}
\label{fig-kpa}
\end{figure}

%%%%%%%%%%%%%%%%%%%%%%%%%%%%%%%%%

These results are not necessarily in contradiction to possible
alignment excesses in other shower observables, at different
observation levels, or in a different primary energy range.
However, the absence of an alignment excess in the present
analysis might be used to constrain attempted alignment
explanations.

The KASCADE data can be compared to the $\lambda_4$ distribution
measured by the PAMIR detector.
PAMIR~\cite{pamir,kop95} is an emulsion chamber
experiment located at an altitude of
4360~m a.s.l.~(corresponding to 600~g~cm$^{-2}$ atmospheric depth),
measuring air shower hadrons and photons with
a particle energy threshold of a few TeV.

In Fig.~\ref{fig-kpa}, the $\lambda_4$ distributions obtained
from KASCADE and PAMIR data~\cite{capsla99,bor97} are compared for
two different primary energies.
In the energy range of a few PeV (upper panel of the
Figure), i.e.\ below the threshold
energy of the alignment excess claimed from PAMIR data, both
distributions and the fraction of aligned events agree well
to each other. This is a quite significant result, since in the
PAMIR experiment, compared to KASCADE,
a completely different detection technique is applied, 
the mass overburden is considerably smaller 
($\Delta X_v \simeq 420$~g~cm$^{-2}$), and the energy threshold of
the registered shower particles is higher by more than an order
of magnitude.
A possible interpretation of this concordance 
is that both $\lambda_4$ distributions are generated mostly from a
random distribution.
Also possible differences in the lateral distributions of the particles
measured by KASCADE and PAMIR
seem too small to significantly affect the $\lambda_4$ distributions. 

This interpretation is supported by comparing the fraction of aligned events
measured by KASCADE $F (\lambda_4 \ge 0.8) \simeq$ (4$\pm$1)\%
with the values derived from RUNJOB of (3$\pm$1)\%~\cite{runjob}
and from NA22 data ($\simeq 6\%$)~\cite{kop95,na22}.
Each result was found to
agree to the respective background fluctuation, and despite the
completely different measurement conditions, the fractions of aligned events
are comparable to each other.

For a higher primary energy range,
based on $\lambda_4$ distributions obtained from PAMIR data
such as shown in Fig.~\ref{fig-kpa} (lower panel),
an alignment excess with a threshold energy of 8$-$10~PeV was 
deduced~\cite{capsla99,bor97}.
Displayed in the Figure are two distributions obtained 
with different reconstruction methods from the same PAMIR data set.
Included in the graph are the KASCADE results for primary energies
above 8~PeV. 
Despite the completely different observation conditions,
the distributions measured by both experiments agree well
to each other.
The fraction of events with $\lambda_4 \ge 0.8$
is larger in the PAMIR data compared to KASCADE.
However, the statistical significance for the fraction of aligned events
to differ between the two experiments is small, amounting to
$\simeq$1.5 standard deviations only.
Thus, also in this primary energy range
the lack of significant differences between the $\lambda_4$ distributions,
including the fraction of aligned events,
suggests a random component 
as a dominant source of the $\lambda_4$ values observed by
both, PAMIR and KASCADE.

The quantity $d_{4}^{\rm max}$, describing distances between
the hadrons as measured by the KASCADE calorimeter, is sensitive
to the transverse momenta in secondary hadron production
and to the primary particle type.
Transverse momenta in high-energy hadron interactions
differing by a factor two or more from the standard values
are disfavoured by the measurements.

%%%%%%%%%%%%%%%%%%%%%%%%%%%%%%%%%%%%%%%%%%%%%%%%%%%%%%%%%%

{\it Acknowledgments.}
The authors would like to thank the members of the engineering and technical
staff of the KASCADE collaboration who contributed with enthusiasm and 
engagement to the success of the experiment.
One of the authors (A.R.) kindly acknowledges helpful discussions
with Prof.~Maria Giller (Lodz) and the kind hospitality at the
Institute of Nuclear Physics PAN (Cracow), where part of the manuscript
was prepared. 
The KASCADE experiment is supported by the German Federal Ministry of
Education and Research (05 CUAVK1/9)
and was embedded in collaborative WTZ projects
between Germany and Poland (POL 99/005), Romania (RUM 97/014) and
Armenia (ARM 02/98).
The Polish group acknowledges the support by KBN grant 
no.~1P03B03926 for the years 2004-2006.

%%%%%%%%%%%%%%%%%%%%%%%%%%%%%%%%%%%%%%%%%%%%%%%%%%%%%%%%%%%

\end{document}